\documentstyle[11pt]{article}

\begin{document}

\font\cmss=cmss10 \font\cmsss=cmss10 at 7pt \hfill CPTH-S493.0197

\hfill hepth/9702056

\hfill January, 1997

\vspace{20pt}

\begin{center}
{\large {\bf CENTRAL FUNCTIONS\\[0pt]
AND THEIR PHYSICAL IMPLICATIONS}}
\end{center}

\vspace{6pt}

\begin{center}
{\sl Damiano Anselmi}

{\it Centre de Physique Theorique, Ecole Polytechnique, F-91128 Palaiseau
Cedex, FRANCE}
\end{center}

\vspace{8pt}

\begin{center}
{\bf Abstract}
\end{center}

\vspace{4pt} I define central functions $c(g)$ and $c^{\prime }(g)$ in 
quantum field theory. These quantities are useful to study the flow
of the numbers of vector, spinor and scalar degrees of freedom from the UV
limit to the IR limit
and justify the notions of secondary
central charges recently introduced
at criticality. Moreover, they
are the basic ingredients for a description of quantum field theory
as an interpolating theory between pairs of four dimensional
conformal field
theories.  
The correlator of 
four stress-energy tensors plays a key role in this respect.
I analyse  the
behaviours of the central functions in QCD, computing their slopes in the UV
critical point. To two-loops, $c(g)$ and $c^{\prime }(g)$ point towards the
expected IR directions. 
As a possible physical application, I argue that a closer study
of the central functions might allow us to lower the upper bound on the
number of generations to the observed value. A similar analysis is carried
out in QED. Finally, 
candidate all-order expressions for the central functions are
compared with the predictions of electric-magnetic duality.

\vfill\eject 

The fundamental properties of conformal field theories in four dimensions
(CFT$_{4}$) are encoded into certain {\sl central charges} $c$, $c^{\prime }$
and anomalous dimensions $h$. The primary central charge $c$ is the Weyl
squared contribution to the trace anomaly. Secondary central charges $%
c^{\prime }$ were defined in ref.\ \cite{noialtri}, motivated by the
observation that the OPE of the stress-energy tensor with itself does not
close, in general \cite{mix}. 
New operators $\Sigma $ are involved, $h$ being their
anomalous dimensions. The first radiative corrections to $c$ and $c^{\prime
} $ were computed in ref.\ \cite{noi} in the context of supersymmetric
theories. The result suggests that the central charges are
invariant under continuous deformations of CFT$_{4}$'s. Computations of $h$
can be found both in \cite{noialtri} and \cite{noi}. Moreover,
in \cite{noi} it was
noted that to the second loop order in perturbation theory, it is
possible, with no additional effort from the computational
point of view, to extend the analysis off-criticality
and study the first terms of quantities that what we can call {\sl central
functions}, which are the main subject of the present paper.

In general, conformal field theories are physically useful to describe the
UV and IR limits of a quantum field theory (when they are well-defined). 
The purpose of the research pursued here is to identify
and analyse the basic quantities that can give us
a description of ordinary quantum field 
theories as theories interpolating between 
pairs of conformal field theories.
The construction of 
natural and, to some extent, unique 
functions $c(g)$ and $%
c^{\prime }(g)\footnote{%
Here $g$ denotes generically the set of coupling constants of the theory.}$
that interpolate between the critical values of the central charges
is nontrivial. Their existence will be proved by a detailed 
renormalization group analysis of the correlator of 
four stress-energy tensors.
In ref.s \cite{noialtri,noi} this issue was not considered and
the present paper is mainly devoted to fill
this gap.
The long-range program is to study the flow 
of the central functions
and, hopefully,
extracting some nontrivial physical information. Precisely, the central
functions $c(g)$ and $c^{\prime }(g)$ allow one to study the flow of the
numbers of vector, spinor and scalar degrees of freedom from the UV
conformal fixed point to the IR conformal fixed point of a generic quantum
field theory.

We recall from the discussion of ref. \cite{noialtri} that
there are three fundamental central charges in CFT$_4$ 
(and, correspondingly, three central functions off-criticality). 
One considers the OPE 
of two stress-tensors and observes that
only the three operators $1$, $\bar{\varphi}\varphi $ and
$J_{\mu }^{5}$ appear before
$T_{\mu\nu}$ itself.
The primary central charge $c$
is associated with the identity operator
and the two secondary charges $c^{\prime
}$ are associated with the mass operator $\bar{\varphi}\varphi $ of the
scalar fields and the axial current $J_{\mu }^{5}$ of the fermions.
These charges are clearly independent and are those that
we call ``fundamental''.
At the moment, we cannot claim that
they
characterize 
CFT$_4$ completely. This issue is an open problem.
Other independent charges
could be generated by the construction that will be presented,
but note that
the next term of the $TT$-OPE is the stress-tensor
itself, which gives back the primary central charge $c$.

The primary central function $c(g)$ is related to the two-point correlator $%
<T_{\mu \nu }(x)\, T_{\rho \sigma }(y)>$, while each one of the secondary
central functions $c^{\prime }(g)$ is related to a channel of the four-point
correlator $<T_{\mu \nu }(x)\, T_{\rho \sigma }(y)\, 
T_{\alpha \beta }(z)\, T_{\gamma
\delta }(w)>$. The reason why the correlators
of the stress-energy tensor are relevant
is that in counting the various kinds of degrees
of freedom it is physically meaningful to use (external) gravity as a
probe. It turns out that it is precisely 
the correlator of four stress tensors that contains the relevant information
for a correct description of CFT$_4$'s and quantum field theories
interpolating among them.

The four dimensional conformal field theories that are most similar to the
two-dimensional ones satisfy the condition $\Sigma =0$ or, equivalently, $%
c^{\prime }=0$, in which case the OPE of the stress-energy tensor closes and
it is possible to define primary fields in the usual way. Unfortunately, 
there is
only one example of this kind: the free vector boson. In the realm of
supersymmetry, on the other hand, 
the OPE's of the stress-energy superfield close only
for the free vector
multiplet. These two simple conformal 
theories are relevant for
QCD and supersymmetric QCD. In the 
infrared limit only the Goldstone bosons survive, so $c$
should flow to zero in absence of quarks. 
Thus, the function $c(g)$ might be
monotonically decreasing from the ultraviolet in pure Yang-Mills theory.
The two-loop radiative correction
to the free value exhibits indeed 
a decreasing behaviour (while $c'$
is identically zero in perturbation theory). The situation is more
intrigueing in presence of quarks, as we shall see. The second part of 
the paper is devoted to the detailed analysis of the slopes of the central
functions in the free critical points of QCD and QED.

In some special models it might be possible to work out all-order
expressions of the central functions, on the same footing, for example, as
the NSVZ beta-function \cite{nsvz} of supersymmetric QCD. In the final part
of this paper some candidate all-order expressions in supersymmetric
theories will be compared with the predictions of electric-magnetic duality 
\cite{emduality}.

The existence of a $c$-theorem \cite{zamolo} in two
dimensions motivated the search for some analogous property
in four dimensions \cite{ctheorem,cappelli}. 
The assumption of a monotonic behaviour of the central functions would
strengthen considerably the physical discussion that we are going to present
here. Nevertheless, just 
proceeding by general physical arguments will suffice to
provide us with an interesting picture.

\phantom{.}

{\it Definition of the first central function.}

The correlator of two stress-energy tensors can always be written in the
form 
\begin{equation}
<T_{\mu \nu }(x)\, T_{\rho \sigma }(0)>=-{\frac{1}{48\pi ^{4}}}X_{\mu \nu \rho
\sigma }\left( {\frac{c(g(x))}{x^{4}}}\right) +\pi _{\mu \nu }\pi
_{\rho \sigma }\left( {\frac{f(\ln x\mu,g(\mu) ))}{x^{4}}}\right) .
\label{decomp}
\end{equation}
where $\pi _{\mu \nu }=(\partial _{\mu }\partial _{\nu }-\delta _{\mu \nu
}\Box )$ and $X_{\mu \nu \rho \sigma }=2\pi _{\mu \nu }\pi _{\rho \sigma
}-3(\pi _{\mu \rho }\pi _{\nu \sigma }+\pi _{\mu \sigma }\pi _{\nu \rho })$. 
$\mu $ is the subtraction point and $g(x)$ is the running coupling
constant. Often, $g(\mu)$ will be simply written as $g$.
$T_{\mu \nu }$ is normalized as $e^{-1}e_{\mu
}^{a}\,\delta S/\delta e_{a}^{\nu }$, $S$ being the action and $e_{\mu }^{a}$
the vielbein.

The argument why only two independent functions can appear in eq.\ (\ref
{decomp}) is the following. The most general expression for $<T_{\mu \nu
}(x)\, T_{\rho \sigma }(0)>$ is the sum of five independent tensors constructed
with $x_{\alpha }$ and $\delta _{\alpha \beta }$. The most general
expression for $<\partial _{\mu }T_{\mu \nu }(x)\,T_{\rho \sigma }(0)>$ is the
sum of three independent terms. So, only two independent functions survive
the imposition of the conservation condition $\partial _{\mu }T_{\mu \nu }=0$.

The function $c$ depends on $x$ only via the gauge coupling $g(x)$,
this being a consequence of the Callan-Symanzik equations and the
finitness of the stress-energy tensor. This property 
is crucial, since a central function should 
depend on the value of the coupling $g$ 
at a {\sl single} scale.
Strictly speaking, in presence of scalar fields $\phi$
the stress-tensor is not 
finite, since the conservation law $\partial_\mu T_{\mu\nu}=0$
allows it to mix 
with the operator $\pi_{\mu\nu}\phi^2$.
Nevertheless, one can immediately 
see that this mixing affects only the function
$f$ and not $c$.
The 
relation between renormalized and bare stress-energy 
tensors is $T^R_{\mu\nu}=T^B_{\mu\nu}+A
\pi_{\mu\nu}\phi^2_B$, $A$ being some appropriate renormalization constant.
The matrix of renormalization constants for the operators 
$({\cal S}_1,{\cal S}_2)\equiv (T_{\mu\nu},\pi_{\mu\nu}\phi^2)$ is then
$\left(\matrix{1& A\cr 0 &
Z_{\phi^2}}\right)$, $Z_{\phi^2}$ being
the renormalization constant of the mass operator.
The Callan-Symanzik equations have to be applied to the set of 
correlators $<{\cal S}_i(x)\,{\cal S}_j(0)>$. Clearly, the
correlators involving $\pi_{\mu\nu}\phi^2$ have the same form as
the second term on the right hand side of eq. (\ref{decomp}) and 
so cannot 
affect the first term.
 
The Callan-Symanzik equations can be
safely applied inside the forth-order differential operators $X_{\mu \nu
\rho \sigma }$ and $\pi _{\mu \nu }\pi _{\rho \sigma }$.
When crossing these operators one has
to worry about possible
ambiguities of the form
$|x|^{4}{\cal P}(x)$ in the functions $c$ and $f$, 
${\cal P}(x)$ being a
polynomial of order less than four (further restricted
by Lorentz invariance to $a+bx^2$).
However, these expressions cannot be generated
in ordinary quantum field theory,
since $c$ and $f$
have a perturbative 
expansion in powers of $\ln x\mu $.

The decomposition (\ref{decomp}) is particularly convenient, because the
natural (but nonlocal) traceless stress energy tensor ${\cal T}_{\mu \nu }$
defined as 
\[
{\cal T}_{\mu \nu }=T_{\mu \nu }+{\frac{1}{3}}{\frac{1}{\Box }}\pi _{\mu \nu
}T, 
\]
($T=T_{\alpha \alpha }$) satisfies 
\begin{equation}
<{\cal T}_{\mu \nu }(x)\, 
{\cal T}_{\rho \sigma }(0)>=-{\frac{1}{48\pi ^{4}}}%
X_{\mu \nu \rho \sigma }\left( {\frac{c(g(x))}{x^{4}}}\right) .
\label{decomp2}
\end{equation}
The
function $c(g)$ appearing in (\ref{decomp}) or (\ref{decomp2})
has the correct properties to be taken as the
all-order definition of primary central function. $c(g)$
correctly interpolates between the values of the central charges at the
critical points. Moreover, it coincides with the definition used in \cite
{noi} to the second loop order in perturbation theory around a free
conformal fixed point.

Regulating the correlator (\ref{decomp}) at $x=0$ and 
taking the derivative with respect to $\ln \mu$, one gets 
a very simple equation for the integrated trace anomaly, namely
a result proportional to
\[
\tilde c(g)\, X_{\mu\nu\rho\sigma}\left(\delta^{(4)}(x)\right)
+{\rm trace \,\, terms.} 
\]
The omitted trace terms can be nonlocal.
With a simple scheme choice, $\tilde c(g)$ is the Borel transform of 
$c(g(x))$ calculated at a special point.
To be precise, let us write $c(g(x))=\sum_n c_n(g)\, t^n$, $t=\ln x\mu$.
Note that $c(g)=c_0(g)$.
$\mu$-independence imposes
$c'_n(g)\beta(g)+(n+1)\, c_{n+1}(g)=0$.
We can regulate according to a formula 
that generalizes the prescription of ref.\ \cite{bible}
\begin{equation}
{ t^n\over x^4}\rightarrow-{n!\over 2^{n+1}}\Box\left(\sum_{k=0}^n {2^k t^{k+1}\over (k+1)!}
{1\over x^2}\right)+a_n\delta(x),
\end{equation}
the constants $a_n$ parametrizing the scheme choice.
With $a_n=0$ one finds $\tilde c(g)=\sum_n n!\, c_n(g)/2^n$. 
A different scheme choice 
just adds a term proportional to the beta-function, which is immaterial
for the computation the IR limit of the central charge. 
For this purpose,
$\tilde c(g)$ is as good as $c(g)$, since the difference between
the two is always proportional to $\beta$. For example, one finds
$c(g)=\tilde c(g)+1/2 \, \beta(g)\, {\rm d}\tilde c(g)/{\rm d}g$
in the scheme $a_n=0$. Off-criticality,  only $c(g)$ is scheme
independent, while $\tilde c(g)$ is not.
At criticality $c(g)=\tilde c(g)$ 
and this is true also at two-loops
around a free theory. 

The existence of a relationship, at criticality, between 
the central charge $c$ and an anomaly in external field 
supports the statement that $c$ is invariant under continuous 
deformations of CFT$_4$'s. Indeed, an anomaly 
in external field is only one loop in a theory 
in which all internal anomalies vanish
(all higher order rescattering graphs resum to zero). 
In ref. \cite{noi} this property was explicitly checked to two-loops,
both for $c$ and $c'$, 
and it was conjectured that
the secondary central charge $c'$
should also be invariant under continuous deformations of CFT$_4$. 
The idea is that $c'$ 
is a generalization of the concept of anomaly in external 
field. Making this statement rigorous
is still an open problem, but the two-loop 
results of \cite{noi} are
a strong support to it.

Note that ${\cal T}_{\mu\nu}$ does not feel the ``improvement ambiguity'' 
$A\pi_{\mu\nu}\phi^2$ typical of scalar fields $\phi$. 
Thus,  the correct general statement for the 
finitness of the 
stress-energy tensor is ${\cal T}_{\mu\nu}^R=
{\cal T}_{\mu\nu}^B$.

Tracing eq. (\ref{decomp}), we get tre two-point trace correlator
\begin{equation}
<T(x)T(0)>=9\Box^2\left( {f(\ln x\mu,g)\over x^4}\right).
\label{*}
\end{equation}

In two dimensions the denominator $x^{4}$ is absent, $X_{\mu \nu \rho \sigma
}=2\pi _{\mu \nu }\pi _{\rho \sigma }-(\pi _{\mu \rho }\pi _{\nu \sigma
}+\pi _{\mu \sigma }\pi _{\nu \rho })$ is identically zero and only the
function $f$ survives. Now, our considerations in higher dimension, in
particular formula (\ref{*}), exclude any role of $f$ at criticality, where
only $c$ should survive, and this raises an apparent puzzle. The solution is
that for a very particular choice, namely $f$ proportional to $\ln x\mu $, $%
\pi _{\mu \nu }\pi _{\rho \sigma }f$ is also traceless \cite{cappelli}. $f=c\ln
x\mu $ is the expression valid at criticality, and $f$ is indeed a good
function interpolating between the critical values of the ultraviolet and
infrared central charges\footnote{$f=c\ln x\mu $ cannot be written as a
function of $g$, of course. This is not the effect 
of improvement terms in the stress-energy tensor (absent in two dimensions). 
Rather, one has to note that in two dimensions the Callan-Symanzik equations
cannot be safely applied inside the forth-order differential 
operator $\pi _{\mu
\nu }\pi _{\rho \sigma }$. One ambiguity, namely the constant
polynomial, survives. As a consequence, the equation $\dot f(\lambda
+t,g)=\dot f(t,g(\lambda ))$ is satisfied 
by the derivative of $f$ with respect to $t=\ln x\mu$ and not by $f$ itself. 
The most general solution is
$f(t,g)=\int_{t_0}^tu(g(t')dt'$ for a certain $u(g)$.}. 
The $c$-theorem \cite{zamolo} relies on this
coincidence of two dimensions.

\phantom{.}

{\it Definition of the other central functions}.

The secondary central functions $c'$ are generated by the correlator
$<{\cal T}_{\mu\nu}(x){\cal T}_{\rho\sigma}(y)
{\cal T}_{\alpha\beta}(z){\cal T}_{\gamma\delta}(w)>$. We first work out
the precise definition of $c'$ and then explain the relationship with
the correlator of four stress tensors. The discussion of the previous
paragraph was just a warm-up for the more involved construction
to be presented now. In particular, we shall see that the primary
central function is just a very special secondary central function.

Generically, we can associate a secondary central function $c_{{\cal O}%
}^{\prime }(g)$ with any operator ${\cal O}$, the most relevant cases being $%
{\cal O=}J_{\mu }^{5}$ and ${\cal O=}\bar{\varphi}\varphi $. In the simplest
situation, one can assume that 
${\cal O}$ does not mix with other operators under
renormalization. More generically, 
the definition that we are going to give applies
to a {\sl set} of operators $\{{\cal O}_{i}\}$ that is closed under
renormalization mixing and irreducible in this sense. For the moment we
assume that the $\{{\cal O}_{i}\}$ have the same canonical dimension $d$.

For example, in ref.\ \cite{noi} the operators $\{{\cal O}_i\}$ are the
various Konishi superfields $K_i$ associated with the different irreducible
representations of the matter multiplets. The $K_i$ mix at the quantum level
and this prevents us from constructing a function $c^{\prime}$ for each of
them: instead, the $K_i$'s contribute to the {\sl same} central function $%
c^{\prime}$, unless some flavour symmetry or similar property allows
one to split the set of Konishi superfields into different irreducible
subsets $\{K_i\}_1,\ldots \{K_i\}_n$, each of which will have its own $%
c^{\prime}$-function $c^{\prime}_1,\ldots c^{\prime}_n$.

In this paragraph we assume that the ${\cal O}_i$ are scalar operators. This
simplifies the formul\ae\ considerably and is sufficient to illustrate how
to prodeed in generality.

Let $Z_{ij}(\ln x^{2}\mu ^{2})$ denote the renormalization constants of the
operators ${\cal O}_{i}$. The correlators 
\begin{equation}
<{\cal O}_{i}(x){\cal O}_{j}(0)>={\frac{1}{x^{2d}}}Z_{ik}(\ln x\mu,g
)A_{kl}(g(x))Z_{jl}(\ln x\mu,g )  \label{k}
\end{equation}
define functions $A_{ij}(g)$ that depend only on $g$. The proof of this fact
is a straightforward application of the Callan-Symanzik equations. On the
other hand, the OPE 
\begin{equation}
{\cal T}_{\mu \nu }(x){\cal T}_{\rho \sigma }(0)\sim {\frac{1}{144\pi ^{2}}}%
X_{\mu \nu \rho \sigma }\left( {\frac{B_{i}(g(x))Z_{ij}^{-1}(\ln
x\mu ,g)}{x^{4-d}}}\right) {\cal O}_{j}(0)  \label{j}
\end{equation}
defines other functions $B_{i}(g)$. We used directly ${\cal T}_{\mu\nu}$
in (\ref{j}) rather than $T_{\mu\nu}$, but an analysis similar
to the one carried out for the primary central function
can be repeated starting from $T_{\mu\nu}$.
We set 
\begin{equation}
c_{{\cal O}}^{\prime }(g)\equiv B_{i}(g)A_{ij}(g)B_{j}(g).  \label{i}
\end{equation}
This function interpolates between the appropriate critical values and
coincides with the function computed in ref. \cite{noi} up to the second
loop order included.

We stress that formul\ae\ (\ref{k}), (\ref{j}) and (\ref{i}) have a
nontrivial meaning also at criticality, if the matter fields are in a
reducible representation of the gauge group. Then, $Z_{ij}=|x\mu
|^{h_{ij}(g)}$, $h_{ij}(g)$ being a matrix that, in general, is not
symmetric. It is simple to work out examples of this kind starting from the
results of \cite{noi}.

We note that $%
c^{\prime}$ is scheme independent, while the quantities $A_{ij}(g)$ and $%
B_{i}(g)$ do depend on the subtraction scheme. To be precise, 
compatibility with the Callan-Symanzik equations requires that 
a scheme change dictated by ${\cal O}_i\rightarrow
H_{ij}(g){\cal O}_j$ be expressed as $Z(\ln x\mu,g)\rightarrow
H(g)Z(\ln x\mu,g)H^{-1}(g(x))$, 
$A(g)\rightarrow H(g)A(g)H^t(g)$, $B(g)\rightarrow
B(g)H^{-1}(g)$. Consequently, 
$c^{\prime}$ is invariant, because the finite subtractions
simplify in expression (\ref{i}): $c'(g)\rightarrow c'(g)$. 
This observation generalizes a remark made
in the context of the two-loop computation of ref.\ \cite{noi} and shows, to
all orders in perturbation theory, off-criticality and in the case of
nontrivial renormalization mixing between operators, that the function $%
c^{\prime }(g)$ is a physically meaningful quantity.

It is straightforward to extend the definition of $c^{\prime }(g)$ to the
case of mixing among operators of different canonical dimensions. This case
is not of primary interest for the applications and is left to the reader.

With a suitable combination of limiting procedures, it is possible to single
out the secondary central functions inside the correlator $<{\cal T}_{\mu
\nu }(x){\cal T}_{\rho \sigma }(y){\cal T}_{\alpha \beta }(z){\cal T}%
_{\gamma \delta }(w)>$. Roughly, one has to
``open'' the four point correlator and look deeply inside.
One first takes the limit in which two distances,
for example $\delta_1\equiv|x-y|$ and $\delta_3\equiv |z-w|$, 
are much smaller than the other one, $\Delta_2\equiv |x-w|$. 
This procedure factorizes two differential operators $X_{\mu
\nu \rho \sigma }$ and $X_{\alpha \beta \gamma \delta }$, acting on a
certain expression ${\cal E}(\delta_1,\Delta_2,\delta_3)=
\sum_{{\cal O}}{\cal E}_{%
{\cal O}}(\delta_1,\Delta_2,\delta_3)$, 
with canonical powers distributed among the
distances $\delta_1$, $\Delta_2$ and $\delta_3$ 
according to the canonical dimension
of the intermediate channel ${\cal O}$. 
Precisely, 
\begin{eqnarray}
\left.<{\cal T}_{\mu
\nu }(x)\, {\cal T}_{\rho \sigma }(y)\, {\cal T}_{\alpha \beta }(z)\,{\cal T}%
_{\gamma \delta }(w)>\right|_{\delta_{1,3}\ll\Delta_2}\sim
{1\over (144\pi^2)^2} \sum_{\cal O}\times
~~~~~~~~~~~~~~~~~~~~~~~~~~~~~~~~~~~~~~~~~~~
\nonumber\\
\overrightarrow{X}^{(1)}_{\mu\nu\rho\sigma}\left(\,
{\frac{B_{i}(g(\delta_1))Z_{ij}^{-1}(\delta_1,g)}{\delta_1^{4-d}}}\,
{\frac{Z_{jk}(\Delta_2,g
)A_{kl}(g(\Delta_2))Z_{ml}(\Delta_2,g )} {\Delta_2^{2d}}}\,
{\frac{Z_{nm}^{-1}(\delta_3,g)B_{n}(g(\delta_3))}{\delta_3^{4-d}}}\,
\right)\overleftarrow{X}^{(3)}_{\alpha\beta\gamma\delta}
\end{eqnarray}
Then, taking the limit  $%
\delta_1=\Delta_2=\delta_3=\lambda $, 
one gets ${\cal E}_{{\cal O}}(\lambda )=c_{%
{\cal O}}^{\prime }(g(\lambda ))/\lambda ^{8}$.

In CFT$_2$, the only channels ${\cal O}$ are the stress tensor
itself and its derivatives\footnotemark\footnotetext{The identity 
operator, on the other hand, cannot be considered as an intermediate 
channel, since it contributes only to disconnected diagrams.}, 
so one gets the usual central charge.
In higher dimensions, instead,
one has many channels ${\cal O}$
and many corresponding central 
charges (or functions, off-criticality)\footnotemark
\footnotetext{Nevertheless, how many of these functions
are independent is still an open problem.
At the moment we can just say that they are at least three.}. 
Consequently, the
basic correlator that one has to consider in order
to properly describe higher dimensional CFT's or 
quantum field theories interpolating among them 
is precisely the correlator of four stress tensors
rather than the correlator of two.

It is true in any dimension that 
the secondary central function associated with
the channel ${\cal O}={\cal T}_{\mu \nu }$ coincides with the primary
central function itself: $c_{{\cal T}_{\mu \nu }}^{\prime }(g)=c(g)$.
This is because $B=1$, $A(g)=c(g)$ in this very special case.  

Before going on, let us summarize the idea behind our construction.
At first, one might think that it is consistent to associate 
a central function to any operator $O(x)$, i.e. the coefficient
of the identity operator in the OPE $O(x)O(0)$.
However, this is not the case. 
First of all, 
one has to eliminate the $Z$-factors, otherwise one does not 
define just a function of $g(x)$, but a function 
of $g(x)$ and $x$ separately: 
see, for example, eq. (\ref{k}). 
But even after getting rid of the $Z$-factors,
$A(g)$ turns out to be a scheme dependent
quantity, as discussed below eq. (\ref{i}).
Moreover, due to operator
mixing, the quantity $A_{ij}(g)$ is in general a matrix
and not a function. 
Finding appropriate $B_i$'s to paste to it while preserving 
a physical meaning is not trivial.
For this reason, just those operators $O_i$ that appear
in the stress-tensor OPE 
are considered here and their appearance in the stress-tensor OPE 
automatically defines 
the appropriate $B_i$'s.
It is clear, then, that the relevant correlator
is the one of four stress-tensors.

Finally, we note that these observations cannot 
be worked out while staying at criticality,
yet they have relevant consequences on CFT$_4$: for example, they show that 
only those quantities that can be promoted to central 
functions off-criticality deserve to be called central charges at criticality.
It is thanks to the construction presented here that
the notion of secondary central charges used for $c'$ in \cite{noialtri,noi}
is firmly justified.

When a quantum field theory admits other
conserved currents $J$ (for example flavour
currents), the construction presented here 
can be generalized in an obvious way to the correlators
$<JJ>$ and $<JJJJ>$. $J$ replaces ${\cal T}$ and defines ``flavour''
primary and secondary central charges.

Having constructed natural central functions $c(g)$ and $c^{\prime }(g)$ 
in a generic quantum field
theory, we are now entitled to study their behaviours in perturbation theory.
We have to expect that the central functions
do not share most of the properties
of the two dimensional $c$-function, apart from interpolating between the
desired UV and IR values. For example, in two dimensions the $c$-function is
stationary (${\rm d}c/{\rm d}\alpha =0$, $\alpha =g^{2}$) at criticality, as
a consequence of the $c$-theorem. This property does not hold in four
dimensions, neither for $c(g)$, nor for $c'(g)$,
as the results of ref.\ \cite{noi} clearly show. So,
the first interesting quantities to be computed 
are the slopes of the central functions
at criticality.
Were we able to compute the central functions of QCD to all orders, we would
presumably be able to extract relevant properties of the infrared limit of
the theory, for example if it confines or not. Moreover, we have given a
rigorous justification of the off-critical analysis of \cite{noi}.

\phantom{.}

{\it Analysis of the slope of the first central function}.

It is not clear whether in QCD the beta function tends to zero or minus
infinity in the infrared limit. If there is no infrared fixed point, then it
might be meaningless to count the degrees of freedom via the central
functions. So, we assume that there is an IR fixed point. Moreover, since
our analysis focusses on a neighborhood of the ultraviolet fixed point, we
can neglect all the quark masses and work effectively with massless QCD.

Let $c^{\prime }$ refer to the
fermions and $c^{\prime \prime }$ to the scalar fields.
In pure Yang-Mills theory, we have $c_{UV}=\frac{1}{10}{\rm \dim }G\equiv 
\frac{1}{10}N_{v}$ and $c_{UV}^{\prime }=c_{UV}^{\prime \prime}=0$. 
We expect $c_{IR}=c_{IR}^{%
\prime }=c_{IR}^{\prime\prime}=0$. 
We must note that $c^{\prime }$ and $c^{\prime\prime}$
are perturbatively zero.
Non-perturbative effects like the glueballs, for example, disappear in the
far infrared because they are massive, but at intermediate energies they are
present. The situation is less simple in presence of quarks, where $%
c_{IR}^{\prime \prime }$ should measure the number of Goldstone bosons, that
are nonperturbative effects surviving in the infrared limit.
With gauge group $SU(N_{c})$ and $N_{q}$
quarks in the fundamental representation, we have $c_{UV}=\frac{1}{10}N_{v}+%
\frac{1}{20}N_{q}N_{c}$, $c_{UV}^{\prime }=N_{q}N_{c}$, $c_{UV}^{\prime
\prime }=0$ and we expect that in the infrared, where only the Goldstone
bosons survive, $c_{IR}=\frac{1}{120}(N_{q}^{2}-1)$ , $c_{IR}^{\prime }=0$
and $c_{IR}^{\prime \prime }=N_{q}^{2}-1$. In QCD we
have $c_{UV}>c_{IR}$ and this inequality holds for a wide range of values of 
$N_{c}$ and $N_{q}$. For example, for $N_{c}=3$ any value $N_{q}$ in the
range of asymptotic freedom is allowed.

The two-loop computation gives 
\begin{eqnarray}
c\left( \alpha _{s}\right) &=&\frac{1}{20}\left[ 2N_{v}+N_{q}N_{c}-\frac{5}{9%
}N_{v}\frac{\alpha _{s}}{\pi }\left( 2N_{c}-{\frac{7}{8}}N_{q}\right)
\right] +{\cal O}(\alpha _{s}^{2}),  \nonumber \\
\left. {\frac{{\rm d}c(\alpha _{s})}{{\rm d}\alpha _{s}}}\right| _{\alpha
_{s}=0} &=&-{\frac{N_{v}}{36\pi }}\left( 2N_{c}-{\frac{7}{8}}N_{q}\right) . 
\nonumber
\end{eqnarray}
This result can be read from \cite{jack} or derived by combining \cite
{hathrell} with \cite{freeman}, after noting that the matter contribution is
the same in QED and QCD to the order considered (apart from the obvious
change in the representations and their Casimirs).

This shows that in pure Yang-Mills theory the central function $c$ does
decrease, the two-loop correction to $c$ having the same sign as the
one-loop beta-function. The presence of matter, instead, produces the
opposite effect, similarly to what happens for the beta function. However,
the effect of matter on the $c$ function is much stronger than the effect on
the beta function, to the extent that an 
inequality like ${\rm d}c(\alpha _{s})/{\rm %
d}\alpha _{s}|_{\alpha _{s}=0}<0$ requires 
\begin{equation}
N_{q}<{\frac{16}{7}}N_{c}.  \label{newc}
\end{equation}
Asymptotic freedom, instead, imposes the less restrictive condition $N_{q}<{%
\frac{11}{2}}N_{c}$. For $G=SU(3)$ (\ref{newc}) becomes noticeably $%
N_{q}\leq 6$. This could be just a coincidence or the sign of some physical
principle, telling us, for example, that once the behaviour of the first
radiative correction off criticality is ``wrong'', then the central function
cannot tend to the appropriate infrared limit. 
Instead, $c(\alpha _{s})$ cannot have a
simple behaviour with $G=SU(3)$ and more than six quark and there
will be 
a value of the coupling constant $\alpha _{s}$ for which ${\rm d}%
c_{s}(\alpha _{s})/{\rm d}\alpha _{s}=0$. 
Pursuing this kind of
investigation further might allow us to lower the upper bound on the number $%
N_{q}/2$ of generations (fixed by asymptotic freedom to be $N_{q}/2\leq 8$)
to the observed value. 

Let us repeat the analysis in QED. We have \cite{hathrell} 
\begin{equation}
c_{e}(\alpha )={\frac{1}{20}}\left( 2+N_{f}+{\frac{35}{36}}{\frac{\alpha }{%
\pi }}N_{f}\right) +{\cal O}(\alpha ^{2}),\;\left. \frac{{\rm d}c(\alpha )}{%
{\rm d}\alpha }\right| _{\alpha =0}={\frac{7}{144}}{\frac{N_{f}}{\pi }>0}.
\label{qed}
\end{equation}
We see that now $c_{e}(\alpha )$ is greater than $c_{e\ free}$. Since the
theory is infrared free, (\ref{qed}) shows that $c_{e}(\alpha )$ decreases
towards the infrared. This behaviour agrees with the one noted in QCD (which
was decreasing from the ultraviolet). There is no constraint, in QED, on the
maximum number $N_{f}$ of charged Dirac fermions, which, on the other hand,
would sound much less natural than an upper bound on the number of
generations.

\phantom{.}

{\it Computation of the second central function to the second loop order in
QCD and analysis of its slope}.

To compute $c^{\prime }$, we use the following trick. A straightforward
analysis of the relevant Feynman diagrams shows that $c^{\prime }$ can only
have the form 
\[
c^{\prime }(\alpha _{s})=N_{q}\,{\rm dim}\,R+y\,N_{q}N_{v}\,C(R)\,{\frac{%
\alpha _{s}}{\pi }}, 
\]
where $R$ denotes the representation of the fermions and $C(R)$ is the
corresponding Casimir. $y$ is an unknown numerical coefficient that we have
to determine. Now, in the particular case of one Majorana fermion ($%
N_{q}=1/2 $) in the adjoint representation, the theory is supersymmetric and 
$c^{\prime }=N_{v}/2+y(N_{v}N_{c}/2)\,\alpha _{s}/\pi $. We observe that in
supersymmetric QCD, as proved in \cite{noialtri,noi}, $c$ and $c^{\prime }$
have to be proportional to each other, since a single superfield, $J_{\alpha 
\dot{\alpha}}$ contains both the stress-energy tensor and the R-current $%
J_{5}$. Since $c=N_{v}/8-(N_{v}N_{c}/32)\,\alpha _{s}/\pi $ \cite{noi}, we
derive $y=-1/4$. We conclude that in QCD we have 
\begin{equation}
c^{\prime }(\alpha _{s})=N_{q}N_{c}-{\frac{1}{8}}N_{q}N_{v}{\frac{\alpha _{s}%
}{\pi }+}{\cal O}(\alpha _{s}^{2}),\;\left. \frac{{\rm d}c^{\prime }(\alpha
_{s})}{{\rm d}\alpha _{s}}\right| _{\alpha _{s}=0}=-{\frac{1}{8}\frac{%
N_{q}N_{v}}{\pi }<0}.  \label{c'}
\end{equation}
In QED this result translates into $c_{e}^{\prime }(\alpha
)=N_{f}-(N_{f}/4)\,\,\alpha /\pi +{\cal O}(\alpha ^{2})$.

We see from (\ref{c'}) that the effective number of fermions decreases from
the ultraviolet, which is good because it should tend to zero in the far
infrared.

We can consider also the quantity 
\begin{equation}
N_v^{eff}\equiv 10c-{\frac{c^{\prime }}{2}}=N_{v}-{\frac{5}{18}}{\frac{%
\alpha }{\pi }}N_{v}\left( 2N_{c}-{\frac{11}{10}}N_{q}\right) .  \label{p}
\end{equation}
Since $c$ counts the total number of degrees of freedom (with appropriate
weights), while $c^{\prime }$ counts the number of fermions, the above
difference should count the effective number of vectors. We see, however,
that in QCD with 6 quarks the above function is slightly increasing from the
ultraviolet, while we would expect it to decrease, since in the far infrared
the vectors should disappear completely. $N_{q}=6$ still plays a special
role, but now 6 is the minimum number of quarks for the above function to 
{\sl increase}.
In general, we see that the functions $c$ and $N_v^{eff}$ are very close to
be constant for $N_{q}=2N_{c}$.
 
\phantom{.}

{\it Candidate all-order expressions and electric-magnetic duality.}

An interesting problem is to work out all-order expressions for the central
functions. This might be very difficult in ordinary theories and could
be easier in supersymmetric theories. In particular, let us consider the
formul\ae\ $c(g,Y)=(3N_{v}+N_{\chi }+N_{v}\beta _{g}/g-\gamma _{i}^{i})/24$
and $c^{\prime }(g,Y)=N_{\chi }+2\gamma _{i}^{i}$ appearing in \cite{noi}.
They hold to two-loops in the most general N=1 supersymmetric theory.
One might ask oneself if these very simple
formul\ae\ are all-order valid\footnote{
The expression of 
anomalous dimension $h$ given in \cite{noi}, on the other hand,
cannot be expected to be all-order valid.}. 
We want now to
compare some consequences of this hypothesis with the predictions of
electric-magnetic duality \cite{emduality}.

We can write (the reader is referred to \cite{noi} for the notation), 
\[
D\equiv 24c+\frac{1}{2}c^{\prime }=3N_{v}+\frac{3}{2}N_{\chi }+N_{v}\frac{%
\beta _{g}(g,Y)}{g}, 
\]
which implies 
\begin{equation}
D_{UV}=D_{IR}.  \label{prediction}
\end{equation}

For an electric theory with $N_{f}$ quarks, $N_{\chi }=2N_{f}N_{c}$. We
shall compare the prediction (\ref{prediction}) with the expectations of
electric magnetic duality in the ``conformal window'' $3/2N_{c}\leq
N_{f}<3N_{c}$, where both the electric and magnetic theories are
asymptotically free and we can compute $D$ in the UV limits. For the
electric theory we have $D_{el}=3(N_{c}^{2}-1+N_{c}N_{f})$. For the magnetic
theory we compute $D_{mag}=3(N_{c}^{2}-1-3N_{f}N_{c}+5N_{f}^{2}/2)$. The two
values do not coincide in general. We conclude that the exactness of the
expressions of $c$ and $c^{\prime }$ of \cite{noi} and electric-magnetic
duality lead to physical predictions that are incompatible.
Hopefully, electric-magnetic duality will teach us how to improve our
two-loop formul\ae.

\phantom{.}

{\it Conclusions}.

We have worked out all-order definitions of certain central functions that
might be useful to extract relevant properties of quantum field theories.
The construction is
non-trivial, but it is also quite general, natural
and, to some extent, unique.
Not all the quantities of CFT$_4$ can be promoted 
to central functions off-criticality and we showed which ones
can and how.

We do not have a method for deriving all-order
expressions of the central functions, for the moment. This task might be very
difficult, but the properties singled out in \cite{noi} and here
make us hope
that it is affordable. In supersymmetric theories, in
particular, non-perturbative ideas might allow us to improve the two-loop
expressions worked out in ref.\ \cite{noi}.

Here we have explored various issues to the second loop order in
perturbation theory. In particular, we computed the slopes of the central
functions in the UV fixed point of massless QCD and in the IR fixed point of
massless QED. We have pointed out some special role played by $N_{q}=6$ for $%
N_{c}=3$ in QCD and, as a possible physical application, we have stressed
that a closer investigation of the central functions might lower the
theoretical upper bound on the number of generations to the observed value.

\phantom{.}

Acknowledgements

I am grateful to U. Aglietti, D.Z. Freedman and A.A. Johansen for useful
correspondence and discussions. This work is partially supported by EEC
grants CHRX-CT93-0340 and TMR-516055.


\begin{thebibliography}{99}
\bibitem{noialtri}  D. Anselmi, M.T. Grisaru and A.A. Johansen, Anomalous
currents, electric-magnetic universality and CFT$_{4}$, HUTP-95/A048,
BRX-TH-388 and hepth/9601023, to appear
in Nucl. Phys. B.

\bibitem{mix} Discussions about 
the nonclosure of the stress-tensor 
OPE's in $d>2$ can be also found in the previous literature.
See, for example, Y. Stanev, Bulgarian Journal of Physics
15 (1988) 93, in $d=4$
and A.C. Petkou, Ann. Phys. 249 (1996) 180 in the $O(N)$ vector
model in $2<d<4$.

\bibitem{noi}  D. Anselmi, D.Z. Freedman, M.T. Grisaru and A.A Johansen,
Universality of the operator product expansions in SCFT$_{4}$, 
Phys. Lett. B324 (1997) 329.

\bibitem{nsvz}  V. Novikov, M.A. Shifman, A.I. Vainshtein, V. Zakharov,
Exact Gell-Mann-Low Function of Supersymmetric Yang-Mills Theories from
Instanton Calculus, Nucl. Phys. B229 (1983) 381;

M.A. Shifman and A.I. Vainshtein, Solution of the Anomaly Puzzle in
supersymmetric Gauge Theories and the Wilson Operator Expansion, {\it ibid.}
B277 (1986) 456.

\bibitem{emduality}  N. Seiberg, Electric-Magnetic Duality in Supersymmetric
Non-Abelian Gauge Theories, Nucl. Phys. B435 (1995) 129.

\bibitem{zamolo}  A.B. Zamolodchikov, ``Irreversibility'' of the Flux of the
Renormalization Group in a 2D Field Theory, JETP Lett. 43 (1986) 730.

\bibitem{ctheorem}  J.L. Cardy, Is There a $c$-Theorem in Four Dimensions?,
Phys. Lett B215 (1988) 749;

H. Osborn, Derivation of a Four-Dimensional $c$-Theorem For Renormalizable
Quantum Field Theories, Phys. Lett. B222 (1989) 97;

I. Jack and H. Osborn, Analogs of the $c$-Theorem for Four-Dimensional
Renormalizable Field Theories, Nucl. Phys. B343 (1990) 647;

G. Shore, A New $c$-Theorem in Four-Dimensions, Phys. Lett. B253 (1991) 380;
The $c_{{\cal F}}$ Theorem,{\it \ ibid.} B256 (1991) 407.

\bibitem{cappelli}  A. Cappelli, D. Friedan and J.I. Latorre, $c$-Theorem
and Spectral Representation, Nucl. Phys. B352 (1991) 616.

\bibitem{bible} D.Z. Freedman. K. Johnson and J.I. Latorre, Differential 
Regularization and Renormalization: a New Method of Calculation in Quantum 
Field Theory, Nucl. Phys. B371 (1992) 353. 

\bibitem{jack}  I. Jack, Background field calculations in curved spacetime
III, Nucl. Phys. B253 (1985) 323.

\bibitem{hathrell}  S.J. Hathrell, Trace anomalies and QED in curved space,
Ann. of Phys. (NY) 142 (1982) 34.

\bibitem{freeman}  M.D. Freeman, Renormalization of Non-Abelian Gauge
Theories in Curved Space-Time, Ann. of Phys. (NY), 153 (1984) 339.

\end{thebibliography}
\end{document}